\documentclass[journal,10pt]{IEEEtran}
\usepackage[utf8]{inputenc} 
\usepackage[T1]{fontenc}    
\usepackage{url}            
\usepackage{booktabs}       
\usepackage{amsfonts}       
\usepackage{nicefrac}       
\usepackage{microtype}      
\usepackage{graphicx}
\usepackage{float}
\usepackage{framed}
\usepackage{mathrsfs}
\usepackage{amsmath}
\usepackage{xcolor}
\usepackage{algorithm}
\usepackage{algorithmic}
\usepackage{bm}
\usepackage{subfigure}
\usepackage{epstopdf}

\title{Dynamic Compression Ratio Selection for Edge Inference Systems with Hard Deadlines}
\author{
    Xiufeng Huang, Sheng Zhou,~\IEEEmembership{Member,~IEEE}
    \thanks{
    This work is sponsored in part by the National Key R\&D Program of China 2018YFB1800804, Nature Science Foundation of China (No. 61871254, No. 91638204, No. 61861136003), and Hitachi Ltd. Part of the paper has been presented in IEEE WCNC 2020 \cite{wcnc2020latency}. (\emph{Corresponding author: Sheng Zhou})

    X. Huang and S. Zhou are with Beijing National Research Center for Information Science and Technology, Department of Electronic Engineering, Tsinghua University, Beijing 100084, China. Emails: huangxf18@mails.tsinghua.edu.cn, sheng.zhou@tsinghua.edu.cn.
    
    }
}

\begin{document}

\maketitle

\begin{abstract}

Implementing machine learning algorithms on Internet of things (IoT) devices has become essential for emerging applications, such as autonomous driving, environment monitoring. But the limitations of computation capability and energy consumption make it difficult to run complex machine learning algorithms on IoT devices, especially when latency deadline exists. One solution is to offload the computation intensive tasks to the edge server. However, the wireless uploading of the raw data is time consuming and may lead to deadline violation. To reduce the communication cost, lossy data compression can be exploited for inference tasks, but may bring more erroneous inference results. In this paper, we propose a dynamic compression ratio selection scheme for edge inference system with hard deadlines. The key idea is to balance the tradeoff between communication cost and inference accuracy. By dynamically selecting the optimal compression ratio with the remaining deadline budgets for queued tasks, more tasks can be timely completed with correct inference under limited communication resources. Furthermore, information augmentation that retransmits less compressed data of task with erroneous inference, is proposed to enhance the accuracy performance. While it is often hard to know the correctness of inference, we use uncertainty to estimate the confidence of the inference, and based on that, jointly optimize the information augmentation and compression ratio selection. Lastly, considering the wireless transmission errors, we further design a retransmission scheme to reduce performance degradation due to packet losses. Simulation results show the performance of the proposed schemes under different deadlines and task arrival rates.

\end{abstract}


\section{Introduction}

The deployment of machine learning algorithms on Internet-of-things (IoT) devices at the network edge \cite{7498684}, can enable many applications such as autonomous driving, intelligent manufacturing, environment monitoring, etc. To address the limited processing capabilities of IoT devices, cloud-based method offloads massive amount of data generated by IoT devices to cloud severs to perform calculation. However, offloading data from the network edge to the cloud sever can bring serious network congestion and long overall latency. Edge computing is proposed to solve this problem \cite{yousefpour2019all}, by exploiting edge servers, which are much closer to the user devices, to process tasks. Nevertheless, the computation capability of edge and the wireless communication resources are still limited. For a resource-constrained edge computing system, resource management is crucial. A light-weight online learning algorithm is proposed based on multi-armed bandit theory to make task offloading decisions among multiple candidate servers, in order to minimize the average task offloading delay \cite{8627987}. On the other hand, some researchers focus on designing resource allocation schemes to minimize the energy consumption or latency by convex optimization \cite{you2016energy}\cite{he2018energy}, Lyapunov optimization \cite{8058414}, reinforcement learning \cite{liu2019resource}\cite{chen2018performance}, etc.

Conventional edge computing considers general computing tasks, largely overlooking the characteristics of specific computing tasks like machine learning. More recently, emerging \textit{edge learning} technology \cite{8736011}\cite{park2018wireless} focuses on deploying machine learning algorithms at the wireless access network edge. Advanced machine learning algorithms, such as Deep Neural Network (DNN), are usually computing and storage intensive tasks. Therefore, the deployment of machine learning algorithms at user devices is difficult due to their limited computation capabilities, storage size and power. To enable DNN at the network edge, some light-weight models such as MobileNet \cite{howard2017mobilenets}, ShuffleNet \cite{zhang2018shufflenet}, are proposed to save computation and memory usages. In addition, model compression technologies, such as weight pruning \cite{han2015learning} and data quantization \cite{oh2018a}, can also support low latency and energy constrained edge inference.

Another promising way to solve the deployment problem is jointly utilizing the computation resources of edge servers and user devices. In co-inference schemes with device-server synergy \cite{li2018edge}\cite{shi2019improving}, user devices finish part of computing tasks and send the intermediate result to edge servers for the rest of computing. A co-inference scheme should allocate the computation tasks among edge servers and user devices, as well as carefully splitting the neural network to reduce the communication cost of sending the intermediate results. For example, Branchynet \cite{teerapittayanon2016branchynet} focuses on neural network splitting in order to provide fast inference while guaranteeing accuracy, and BottleNet++ \cite{Shao2019BottleNetAE} adds a pair of encoder-decoder on the split point to perform joint source-channel coding. However, these methods fail when the user devices do not have the model, for example when the model is continuously trained on the edge server and broadcasting the model is too costly.

We therefore consider an edge learning system where user devices send data to the edge server for inference. In this case, new challenges emerge: the unreliable wireless transmission of raw data, typically with large size, can easily violate the task latency deadlines. To optimize the edge inference performance, in terms of accuracy and task completion latency, we exploit a major characteristic of machine learning: the \emph{information redundancy} in the input data \cite{zurada1997perturbation}. Removing redundant information can save communication resources, while at the time may not significantly affect the inference accuracy. This inspires us to use lossy compression before transmission. Energy-efficient lossy compression algorithm is exploited for IoT devices to perform edge inference \cite{AZAR2019168}. Ref. \cite{8635566} investigates saving energy by joint data compression, computation offloading and resource allocation. However, these works mainly aim at the energy saving issue, rather than the inference performance (e.g. the inference accuracy, latency, and etc) of the machine learning model. If the compression ratio, defined as the ratio of the size of the raw data to the size of compressed data, is high, the data can be transmitted faster but the inference accuracy will degrade. To keep certain accuracy, additional retransmissions may occur and inevitably bring excessive latency. If the compression ratio is low, the learning model can perform better at the cost of long communication latency. In short, the selection of the compression ratio should balance the tradeoff between the transmission latency and inference accuracy.


Realizing that different tasks have different information redundancy in the raw data, intuitively one should select higher compression ratio for the tasks with more redundancy. However, calculating the information redundancy before transmission and inference is impractical, and thus we propose an information augmentation scheme. The user devices can first transmit data with high compression ratio, with only few communication resources. If the inference result is wrong, the user devices can perform information augmentation by retransmitting the less compressed task. The challenge is that the edge inference system may not be able to determine whether the result is correct. Luckily, \emph{uncertainty} from the machine learning output can be exploited to estimate the confidence of the inference result \cite{finlay2019empirical}.

Finally, because the wireless channel is unreliable, the inference task can fail due to packet losses. Retransmission \cite{lin1984automatic-repeat-request} can solve this problem but results in extra delay. Due to the extra delay, task may break the latency deadline and even the transmissions of following tasks are affected. To this end, the compression ratio selection should jointly consider the packet loss probability, the queue state of waiting tasks, and the inference accuracy with different compression ratios.

In this paper, we consider an edge inference system with random arrival of tasks, where tasks are offloaded to the edge server for inference. To maximize the number of tasks with correct inference results subject to hard latency deadlines, we use lossy compression on the raw data and design a dynamic compression ratio selection scheme to balance the tradeoff between transmission latency and the inference accuracy. Furthermore, we propose information augmentation to save more communication resources, and retransmission scheme, trying to avoid the performance degradation due to packet losses. In particular, our contributions include:

 \begin{itemize}
    \item Lossy compression of raw data before transmission is exploited to save communication resources in a latency guaranteed edge inference system. By modeling the relation between transmission delay and inference accuracy under lossy compression, we formulate the design of dynamic compression ratio selection as an online optimization problem with stochastic task arrivals, in order to balance the tradeoff between transmission delay and inference accuracy, under the hard latency deadlines.

    \item To solve the proposed problem of dynamic compression ratio selection, we first design an offline algorithm using dynamic programming (DP) to obtain the performance upper bound. We then suppose that the arrival process of tasks is a Bernoulli process, and propose an online algorithm using Markov decision process (MDP). Experiments show that the online algorithms can perform almost the same as the offline one.

    \item We further propose an information augmentation scheme to spend fewer communication resources on the tasks with more information redundancy. Edge devices can transmit data with high compression ratio at first and retransmit it with lower compression ratio if the inference result is wrong. However, it is usually difficult to judge whether the result is correct or not, therefore we exploit the concept of uncertainty in machine learning to estimate the confidence of the inference result. The challenge of applying MDP to jointly determine the compression ratio and additional transmissions is that, the number of transmissions of each task is a random variable, and thus the number of state transition related to one task is uncertain. Therefore, we design a new state transition policy of MDP, by making every optional compression ratio correspond to one state transition even if it is not selected, to ensure that every task has the same number of state transitions. Experiments show that information augmentation can bring performance improvement, especially when the arrival rate or the latency requirement is stringent.

    \item We address transmission errors in the wireless channel by retransmission for packets that are lost, and the key is to determine whether retransmission is needed and the compression ratio used for retransmission. Experiments show that the performance degradation is less than 2\% even when the probability of packet loss reaches 10\%, compared with the case of no packet loss.
 \end{itemize}

The rest of this paper is organized as follows. In Section \ref{sec:sysmod} we introduce the system model. In Section \ref{sec:scheme} we introduce the proposed transmission schemes in details, including offline algorithm, online algorithm, information augmentation scheme and retransmission scheme. Experiments results are provided in Section \ref{sec:exp}. The paper is concluded in Section \ref{sec:conclusion}.

\section{System Model}\label{sec:sysmod}

\begin{figure}[t]
    \centering
    \includegraphics[width=1.1\linewidth]{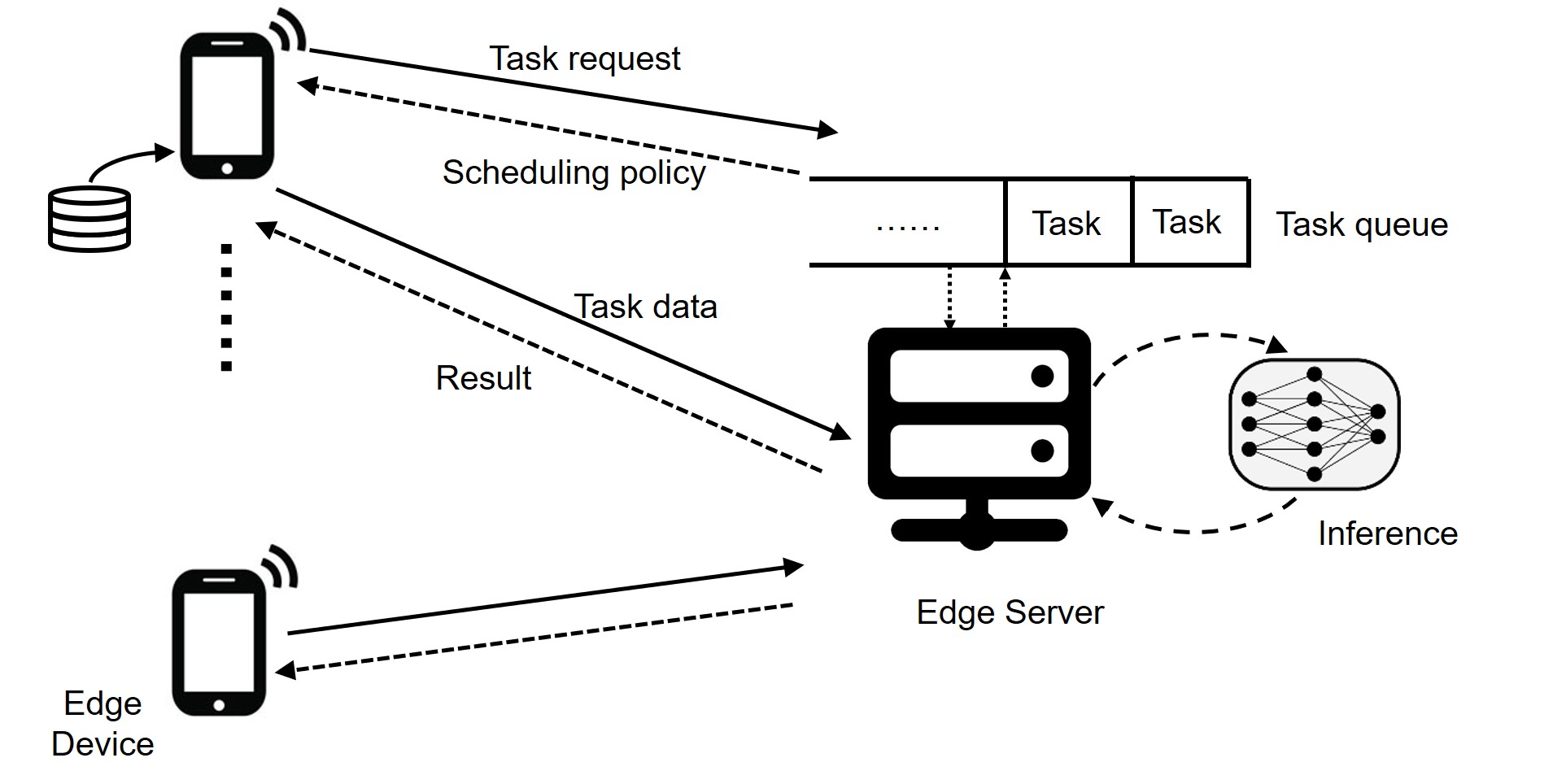}
    \caption{The edge learning system under consideration.}
    \label{fig:system}
\end{figure}

Fig.\ref{fig:system} shows the edge learning system under consideration, which consists of an edge server (attached to a base station) and several edge devices. At the edge server, there is a well-trained learning model used for processing tasks.

The system is time slotted. There are random task arrivals at every edge device. When a task arrives at a device, the device sends a request to the edge server for performing calculation, i.e., inference, and waits for being scheduled by the server, in order to transmit the task data. All tasks have the same maximum latency deadline $\tau$, meaning that every task should be completed before $\tau$ time slots after arriving at the device, otherwise the task is considered failed. The time used for processing tasks at the server side, and sending requests as well as scheduling decisions is the same for different tasks. Therefore, it can be considered by subtracting the total latency constraint with the corresponding overhead, i.e., with larger time consumption of task processing, sending requests and scheduling decisions, the system has smaller $\tau$ for communications. The task requests from all edge devices form a task queue on the edge server, as all requests are recorded. The arrival process of the task queue is assumed to be a Bernoulli process, meaning that there is at most one task arriving at the edge server in every time slot with probability $p$. As all tasks have the same priority, the edge server adopts the First Come First Serve (FCFS) scheduling principle. With FCFS, the tasks from different devices have the same probability to be served, and thus the system can ensure the fairness among tasks in the statistical sense. As a result, even if edge devices have different task arrival rates, the edge server will perform inference for the same proportion of tasks from each device.

Transmitting raw data for these tasks is time consuming. Therefore, lossy compression is used to reduce the transmission latency. However, the compression of data can degrade the inference performance. To balance the tradeoff between the transmission time and the inference accuracy, the edge server should select the appropriate compression ratio for every task according to the state of the task queue. The inference accuracy can be regarded as the reward of performing inference. In the considered system, the edge server performs inference for the same class of tasks, whose data is collected by the same category of sensors or devices. Therefore, we assume that all tasks have the same raw data size and the reward is a function of compression ratio. In the scenarios that tasks have different raw data sizes, we can use the reward as a function of the data size being used, to reflect both the impact from the raw data size and the compression. Nevertheless, adapting the reward function does not change the nature of our proposed algorithm, and thus we stick to the assumption of uniform raw data size. Compression ratio $r \in [1,+\infty)$ is defined as the ratio of the size of the raw data to the size of compressed data, i.e., larger compression ratio leads to less data size for transmission. The accuracy of the machine learning model with compression ratio $r$, i.e. the reward function, is $\rho(r) \in [0,1]$, which can be obtained offline by performing inference on a validation dataset and stored as a lookup table at the server side. The edge devices use fixed rate for transmission and the time slots used for transmission using compression ratio $r$ is $T(r) \in \mathbb{N}$. $\rho(r)$ and $T(r)$ are both decreasing functions. The wireless channel is assumed to be block fading, that is, the fading is i.i.d. among different slots. In every time slot, the scheduled edge device transmits a packet of data samples, which will be lost with probability $p_e$. As a result, transmission of task using $T(r)$ time slots will fail with probability
\begin{equation}
P_e(T(r)) = 1 - \left(1-p_e\right)^{T(r)},
\end{equation}
which is also called as packet error ratio (PER).

\begin{figure}[ht]
    \centering
    \includegraphics[width=\linewidth]{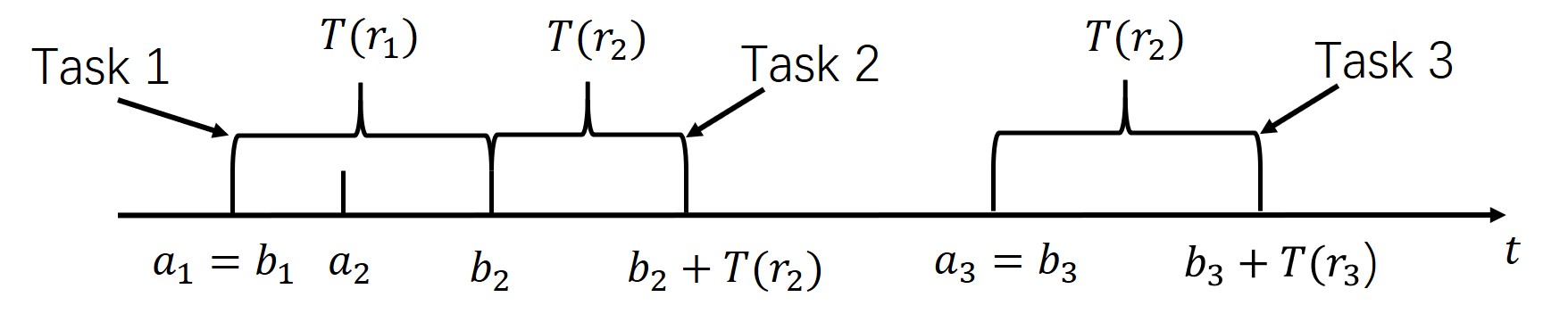}
    \caption{Example of the arriving and transmitting of tasks.}
    \label{fig:time}
\end{figure}

The objective of the transmission scheme is to maximize the expectation of the number of successfully completed tasks with the deadline $\tau$ as shown in (\ref{eq:problem}), considering $M$ tasks, whose arrival time is $a_1 < a_2 < ... < a_M$. The $i$-th task is scheduled to transmit data from time slot $b_i$ with compression ratio $r_i$. Fig. \ref{fig:time} shows an example with $3$ tasks.

The optimization problem is formulated as
\begin{align}
&&\max\limits_{r_i}\ &\sum\limits_{i=1}^M \rho(r_i)\left(1-P_e\left(T(r_i)\right)\right)\label{eq:problem} \\
&&\text{s.t.}\ &b_1 = a_1,\label{eq:b1} \\
&&\ &b_{i+1} = \max \{b_i + T(r_i), a_{i+1} \},  i=1,2,\ldots,M-1,\label{eq:bi} \\
&&\ &b_i + T(r_i)\leq a_i + \tau,  i=1,2,\ldots,M, \label{eq:latency}\\
\nonumber
\end{align}
where (\ref{eq:b1}) and (\ref{eq:bi}) indicate that the $(i+1)$-th task will be transmitted when the transmission of the $i$-th task is completed (if the $(i+1)$-th task has not arrived, then it will be transmitted when it arrives). Equation (\ref{eq:latency}) is the latency deadline constraint. Notice that $T(r_i)$ can equal to $0$ and then $\rho(r_i)$ will also equal to $0$ (it means that the task is failed since it cannot be delivered before the deadline), which ensures that (\ref{eq:latency}) can be satisfied for all tasks.

\section{Proposed Transmission Schemes}\label{sec:scheme}

\subsection{Offline Algorithm}

An offline algorithm can solve the optimization problem described in section \ref{sec:sysmod} if the arrival time of all the tasks is known beforehand. The proposed offline algorithm is designed using DP. We use $F(m,t),(1 \le m \le M, 1 \le t \le a_M + \tau)$ to denote the maximum number of successfully \emph{completed} tasks when processing the first $m$ tasks by time $t$. To get $F(m,t)$, the number of time slots used for transmitting the $m$-th task is denoted by $G(m,t)$. Then we can get the optimal offline policy, i.e., the selected compression ratio for every task $r(m)$, after calculating $F(m,t)$ and $G(m,t)$ ($1 \le m \le M, 1 \le t \le a_M + \tau$). The algorithm is shown as Algorithm \ref{alg:offline}.

\begin{algorithm}
    \caption{Offline DP algorithm}
    \label{alg:offline}
    \begin{algorithmic}[1]
    \REQUIRE ~~\\
        Number of tasks $M$;\\
        Maximum waiting time of tasks $\tau$;\\
        Arriving time of tasks $a_m$;
    \ENSURE ~~\\
        Maximum number of successfully completed task $F(m,t)$;\\
        Optimal policy $r(m)$;
    \STATE {set $F(0,t)=0\ (1\le t \le a_M+\tau)$}
    \FOR{$m=1$ to $M$}
        \FOR{$t=a_m+1$ to $a_m+\tau$}
            \STATE \begin{align}
            F(m,t) = &\max \limits_{a_m\le i \le t-1} \bigg(F(m-1,i)+ \nonumber \\
            &\rho\Big(T^{-1}(t-i)\Big)\Big(1-P_e(t-i)\Big)\bigg)\nonumber
            \end{align}
            \STATE \begin{align}
            G(m,t) = &\mathop{\arg\max}\limits_{a_m \le i \le t-1} \bigg(F(m-1,i)+ \nonumber \\
            &\rho\Big(T^{-1}(t-i)\Big)\Big(1-P_e(t-i)\Big)\bigg)\nonumber
            \end{align}
        \ENDFOR
    \ENDFOR

    \STATE set $t = a_M + \tau$
    \STATE set $m = M$
    \WHILE {$m>0$}
        \STATE $r(m) = T^{-1}(G(m,t))$
        \STATE $t = \min(t-G(m,t), a_{m-1}+\tau)$
        \STATE $m = m-1$
    \ENDWHILE
    \end{algorithmic}
\end{algorithm}

\subsection{Online Algorithm}\label{sec:online}

In practice, the arrival times of tasks cannot be known beforehand. Therefore an online algorithm is needed to select the compression ratios for the tasks in the queue without the knowledge of future arrivals. Assume that the arrival process is known, which is a Bernoulli process with probability $p$. Then, we can use MDP to solve this problem, by introducing the state, action, reward and state transition probability  of MDP as follows. With MDP, we can map the task queue to the MDP state and use the optimal action based on the state to select the optimal compression ratio for the head-of-line (HoL) task.

\textbf{State}: To make the decision on the compression ratio $r$ for the HoL task in the current time slot, the information needed is the arrival time of all tasks waiting in the queue, which can be represented by $s=\{a_1,a_2,...,a_N\}$, where $N$ is the number of tasks in the waiting queue and $a_1 < a_2 < ... < a_N$. Due to the deadline constraint, $N$ is at most $\tau$.  Note that the difference between the deadline of tasks and current time is sufficient for making decisions, the state is transformed to $s=\{a_1+\tau-t,a_2+\tau-t,...,a_N+\tau-t\}$, where $\tau$ is the latency constraint of tasks and $t$ is the current time. Because $0 < a_1+\tau-t < a_2+\tau-t<...<a_N+\tau-t \le \tau $, the state $s$ can be encoded to a binary number of $\tau$ digits
\begin{equation}
 s=\sum\limits_{i=1}^N 2^{a_i+\tau-t-1}.
\end{equation}
In the binary expression of $s$, the number of 1's equals to the number of tasks in the queue and the positions of 1's represent the remaining time of tasks (lower digit represents fewer remaining time slots).

\textbf{Action}: The action of the MDP is the optional compression ratio of the HoL task in the queue. The action space is defined as $\mathcal{R}$, which is the set of optional compression ratios of the edge inference system. Furthermore, in the case of using the reward as a function of data size, the system should include the raw data sizes of tasks into the state of MDP, and use the data size, which should be smaller than the raw data size, as the action of MDP, rather than using the compression ratio.

\textbf{Reward}: The reward of taking action $r \in \mathcal{R}$ (selecting compression ratio $r$) is
\begin{equation}
W(r) = \rho(r)[1-P_e(T(r))],
\end{equation}
which is the product of the expected accuracy of the machine learning model with compression ratio $r$ and the probability of successful transmission.

\textbf{State transition probability}: The state transition probability depends on the action and the arrival process of tasks. With action $r$, state $s$ will transit to state $s' = 2^{\tau-T(r)}i + \left\lfloor \frac{s}{2^{T(r)}} \right\rfloor\ (i=0,\ 1,\ ...,\ 2^{T(r)}-1)$ with probability
\begin{equation}
\mathbf{P}_{ss'}(r) = p^{B(i)}(1-p)^{T(r)-B(i)},
\end{equation}
where $B(i)$ is the number of 1's in the binary expression of $i$.

Generally, one needs to calculate the state transition probability matrix $\mathbf{P}$ and perform value iteration, whose space complexity and time complexity are both $\mathcal{O}(S \times S \times |\mathcal{R}|)$ where $S=2^{\tau}-1$ is the number of states. The storage of the state transition probability matrix is unacceptable if $\tau$ is large and calculating the state transition probability when performing value iteration is not effective. However, using the similarity of transition probability of different states and the fact that there are many zero elements in the state transition probability matrix, we do not need to calculate and store the state transition probability matrix explicitly, then we can greatly reduce the complexity of the algorithm as follows. First, the value iteration equation of MDP is
\begin{equation}\label{eq:vl1}
V^{k+1}(s) = \min \limits_{r \in \mathcal{R}(s)} \left[W(r) + \sum\limits_{i=1}^{2^{\tau}-1} \mathbf{P}_{si}(r)V^k(i)\right],
\end{equation}
where $V^k(i)$ is the value of state $i$ of the $k$-th iteration and $R(s)$ is the set of optional action of state $s$. To reduce the complexity, we use (\ref{eq:vl2}) for value iteration.
\begin{equation}\label{eq:vl2}
\begin{aligned}
V^{k+1}(s) = &\min\limits_{r \in \mathcal{R}(s)}\Bigg[W(r) + \\
             &\sum\limits_{i=0}^{2^{T(r)}-1} \mathbf{P}'(r,i)V^k\left(2^{\tau-T(r)}i + \left\lfloor \frac{s}{2^{T(r)}} \right\rfloor\right)\Bigg].
\end{aligned}
\end{equation}
For (\ref{eq:vl2}), the state transition probability is $\mathbf{P}'(r,i) = p^{B(i)}(1-p)^{T(r)-B(i)}$. For efficient value iteration, $\mathbf{P}'$ is calculated and stored before performing value iteration. Therefore we only need to store $\mathbf{P}'$, with $S \times |\mathcal{R}|$ elements, instead of $\mathbf{P}$. And equation (\ref{eq:vl2}) only sums up $2^{T(r)}-1$ terms when considering action $r$, instead of $2^{\tau}-1$ terms in (\ref{eq:vl1}). With the optimization in (\ref{eq:vl2}), the space complexity is now $\mathcal{O}(S \times |\mathcal{R}|)$ and the time complexity is $\mathcal{O}(S \times S \times |\mathcal{R}| \times \frac{1}{\tau})$.

It is shown that the complexity of the proposed algorithm highly depends on $\tau$. In Section \ref{sec:sysmod}, we assume that in every time slot there is at most one task arriving at the edge server. When the number of devices in the system and the task arrival rate at the edge server become larger, we need to use shorter time slot to satisfy the assumption, which will result in larger $\tau$ (in terms of the number of time slots). In this case the edge inference system may fail to make scheduling decisions due to the large complexity. However, to have reasonable performance of an edge learning system with finite computation capability, the number of active devices should be limited, i.e., the task arrival probability $p$ cannot be too large. In many IoT systems, even if there are many devices, their active probability is still low. In short, as long as the computation load is reasonable, the algorithm can scale with corresponding $p$ and $\tau$.

\subsection{Information Augmentation Scheme}

\begin{figure}[t]
    \centering
    \includegraphics[width=1.05\linewidth]{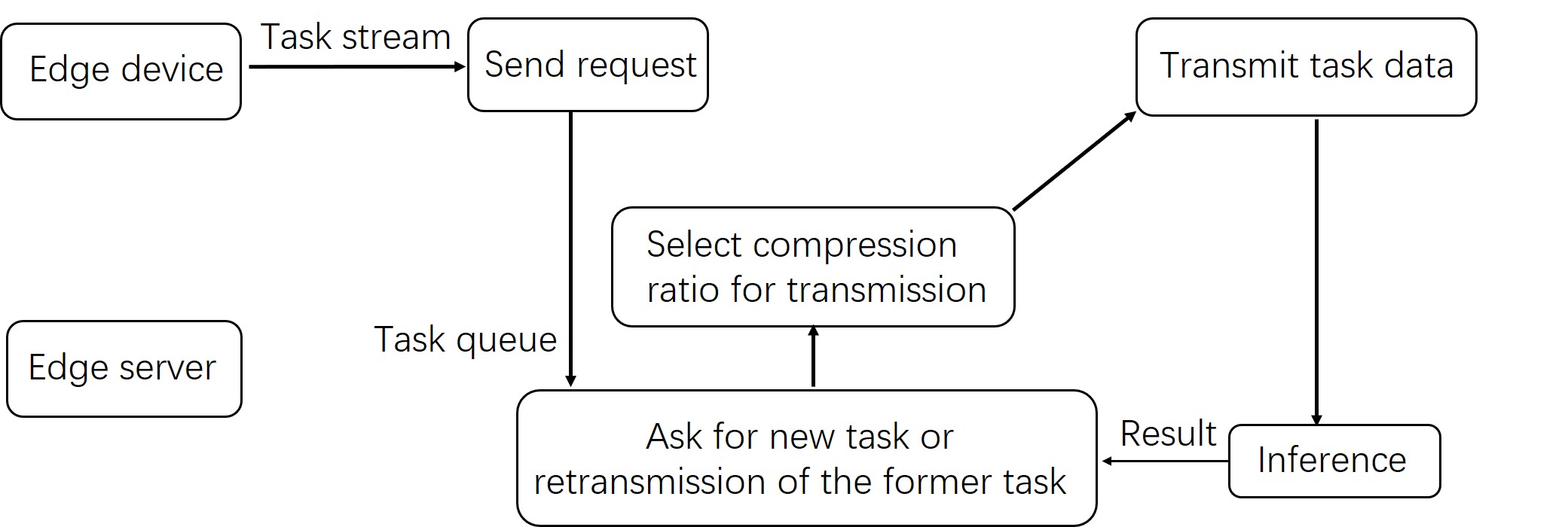}
    \caption{Workflow of information augmentation scheme.}
    \label{fig:retrans}
\end{figure}

For machine learning algorithms, some tasks can get correct inference result with very high compression ratio, because of many information redundancies in the raw data. If we spend fewer communication resources on these tasks, left more communication resources for tasks that require lower compression ratio, we can complete more tasks with limited communication resources. Therefore, we propose an information augmentation scheme to find the proper compression ratio for task and the workflow is shown in Fig. \ref{fig:retrans}.

We first assume that there is a method to judge whether the result is correct or not and the packet loss probability $p_e=0$. As shown in Fig. \ref{fig:retrans}, with the proposed information augmentation scheme, the edge server can first ask the edge devices to transmit data with a high compression ratio (not necessarily the highest one, and the compression ratio for the first attempt is also subject to our optimization) . If the result is wrong, the edge server can decide to ask the device to transmit the task data with lower compression ratio, according to the state of task queue. In short, information augmentation scheme needs to exploit the state of the task queue to decide the compression ratio of the first transmission attempt and later on augmentation transmissions if needed.

We also use MDP to solve the decision making problem in the proposed information augmentation scheme. In the last subsection, we use state $s=\{a_1,a_2,...,a_N\}$ for MDP for the transmission scheme without information augmentation and the action space $R$ is the set of optional compression ratios. Notice that one state transition corresponds to one transmission and the objective of MDP is to maximize the average reward per state transition. However, for the information augmentation scheme, the times of transmission of different tasks may be different. If one state transition of MDP still corresponds to one transmission over the wireless channel, different tasks will have different weights of the reward, and thus we cannot directly apply the MDP formulation from the last subsection.

To ensure that different tasks have the same times of state transitions, we consider a new state space and new state transition formulation. For every task, the edge server will virtually \emph{consider} all optional compression ratios from high to low for its transmission (if the task can be completed with a high compression ratio, the transmission with lower compression is not needed). Every state transition corresponds to one time of \emph{virtual consideration}. There are only two actions for MDP, transmission (or retransmission) with the current compression ratio or no transmission. If the action is no transmission, the edge server will consider the next compression ratio. If the action is to transmit, then the task will be transmitted with the current compression ratio under consideration. With the inference result and the state of task queue, the edge server will continue to consider the next compression ratio for possible information augmentation, even if they are not selected. In this way, every task will have $|R|$ times of state transitions. The state, action, reward and state transition probability of the MDP of the proposed information augmentation scheme is shown as follows.

\textbf{State}: The key here is that we put the compression ratio under consideration into the state, and now the state is $s=\{a_1,a_2,...,a_N,r_L,r,f\}$, where $r_L$ is the compression ratio for the last transmission for current task (if the task has not been transmitted, $r_L=+\infty$), $r$ is the compression ratio considered for current transmission and $f \in \{0,1\}$ is the correctness of the result of the last transmission ($1$ represents correct and $0$ represents wrong). For simplicity, as in the last subsection, the state can be rewritten as
\begin{equation}
s=\{\sum\limits_{i=1}^N 2^{a_i+\tau-t-1},r_L,r,f\}=\{a,r_L,r,f\},
\end{equation}
where $a=\sum\limits_{i=1}^N 2^{a_i+\tau-t-1}$ represents the state of the task queue.

\textbf{Action}: Because the compression ratio under consideration is formulated into states, there are only two actions for MDP, transmit or not (denoted by $r_I$, $r_I=1$ represents transmission and $r_I=0$ represents no transmission). If the result of last transmission is correct ($f=1$) or the rest latency budget is not enough for additional transmission, there is only one choice of no transmission.

\textbf{Reward}: For the state transition from $s=\{a,r_L,r,f\}$ to $s'=\{a',r'_L,r',f'\}$ with action $r_I$, the reward is
\begin{equation}
W_I(s,s',r_I) = r_If'.
\end{equation}

\textbf{State transition probability}: The state transition probability depends on the arrival process of tasks and the accuracy of inference conditioned on the inference correctness of the last transmission. If the action is no transmission, then the state $s=\{a,r_L,r,f\}$ will transit to state $s'=\{a,r_L,r',f\}$ with probability $1$, where $r'$ is the next considered compression ratio after $r$. If the action is to transmit, then $s=\{a,r_L,r,f\}$ will transit to state $s'=\{\lfloor \frac{a}{2^{T(r)}} \rfloor + i\times2^{\tau-T(r)},r,r',f\}\ (i \in [0,2^{T(r)}-1] \cap \mathbb{N})$ and the state transition probability is
\begin{equation}\label{eq:retansP}
\mathbf{P}(s,s') = p^{B(i)}(1-p)^{T(r)-B(i)} \times \mathbf{P}_a(r,f|r_L).
\end{equation}
$\mathbf{P}_a(r,0|r_L)$ is the probability that the inference result is wrong with compression ratio $r$ conditioned on that the last transmission was with compression ratio $r_L$ and the inference result was wrong (otherwise there is no need to transmit again). $\mathbf{P}_a(r,1|r_L)$ is the probability that the inference result is correct with compression ratio $r$ conditioned on that the last transmission was with compression ratio $r_L$. $\mathbf{P}_a$ is obtained by performing inference on the validation dataset with optional compression ratios.

The number of states is $S_R = 2(2^{\tau}-1)|R|^2$ and the size of action space is $2$. The memory usage is the simplified state transition matrix mentioned in last subsection and $\mathbf{P}_a$. As a result, the space complexity is $\mathcal{O}(S_R)$ and the time complexity is $\mathcal{O}(S_R \times S \times \frac{1}{\tau})$.

\subsection{Information Augmentation with Uncertainty}

\begin{figure*}[ht]
\centering
    \subfigure[$4 \times 4$]{
    \begin{minipage}[t]{0.25\linewidth}
        \centering
        \includegraphics[width=\linewidth]{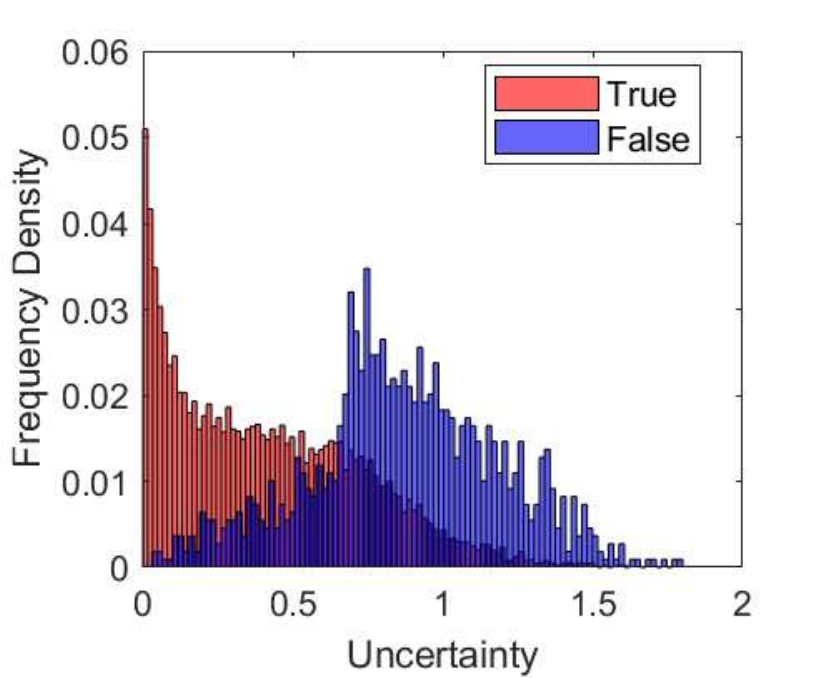}
    \end{minipage}%
    }%
    \subfigure[$7 \times 7$]{
    \begin{minipage}[t]{0.25\linewidth}
        \centering
        \includegraphics[width=\linewidth]{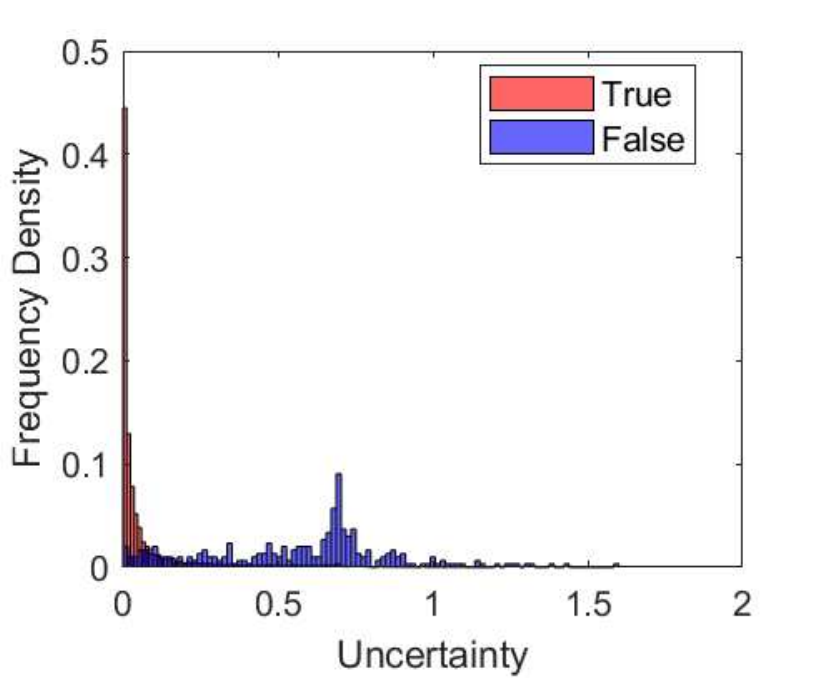}
    \end{minipage}%
    }%
    \subfigure[$14 \times 14$]{
    \begin{minipage}[t]{0.25\linewidth}
        \centering
        \includegraphics[width=\linewidth]{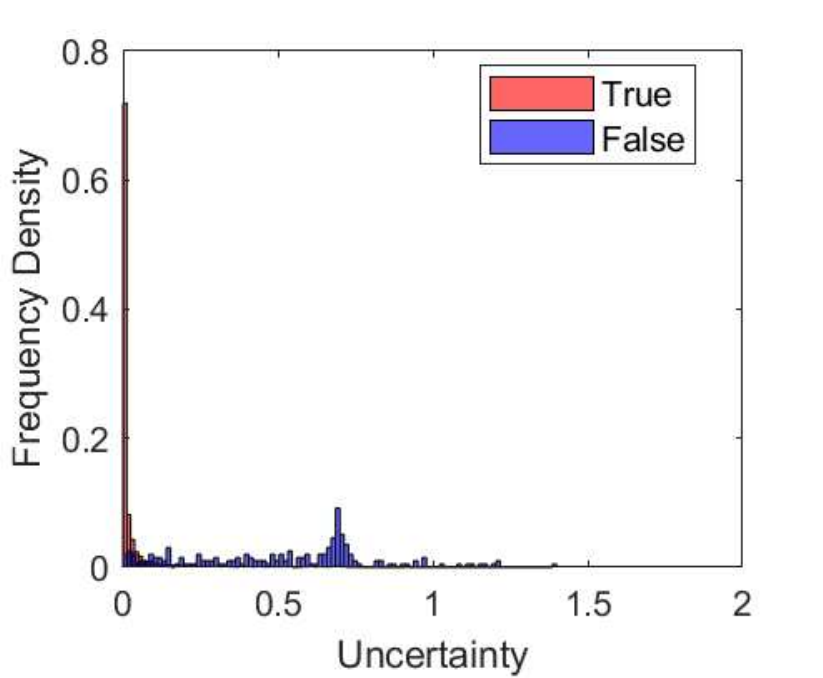}
    \end{minipage}
    }%
    \subfigure[$21 \times 21$]{
    \begin{minipage}[t]{0.25\linewidth}
        \centering
        \includegraphics[width=\linewidth]{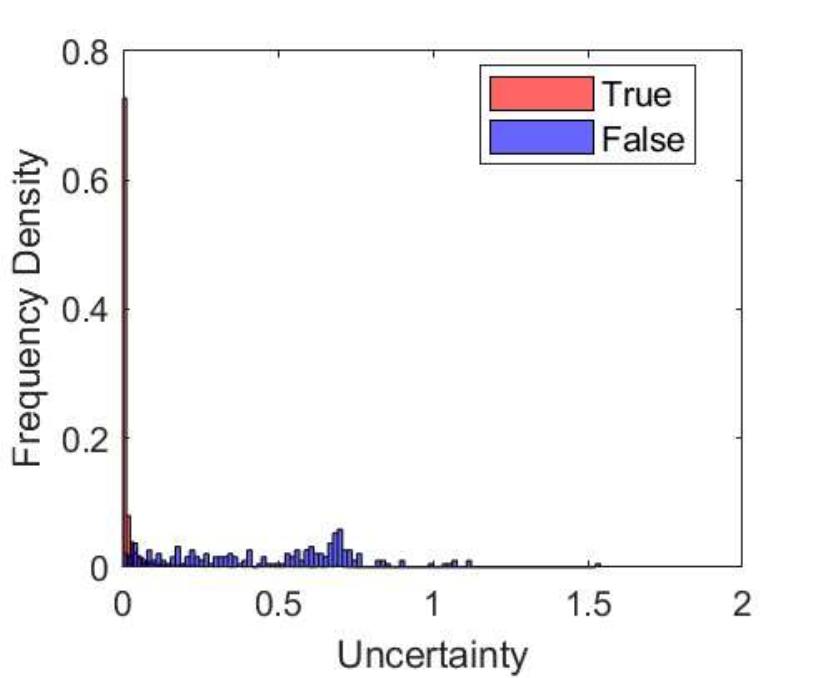}
    \end{minipage}
    }%
\centering
\caption{Histogram of uncertainty.}
\label{fig:hist}
\end{figure*}

In the last subsection, we assume that there is a method to judge whether the inference result is correct or not. However, this assumption is invalid in many scenarios. To solve this problem, we further introduce uncertainty to estimate the confidence of the inference result. The uncertainty $\mathcal{U}$ of the output of the learning model is defined in (\ref{eq:uncertainty}),
\begin{equation}\label{eq:uncertainty}
\mathcal{U} = -\sum\limits_{i=1}^{n} X_i\log X_i,
\end{equation}
where ${\bm X} = (X_1,X_2,...,X_n)$ is the normalized output of learning model ($\sum\limits_{i=1}^n X_i = 1$) and $n$ is number of elements of the model output. The lower uncertainty $\mathcal{U}$ indicates higher probability of correct inference \cite{finlay2019empirical}. For example, the relation between correctness and uncertainty of data samples in MNIST \cite{726791}, a handwritten digit images dataset, is shown in Fig. \ref{fig:hist}. It shows the histogram of uncertainty of correct results and wrong results. Different subfigures correspond to different resolutions. It is shown that the true results have lower uncertainty and the wrong results have higher uncertainty. Therefore it is reasonable to exploit uncertainty for estimating the confidence of the inference results.

The problem of selecting the compression ratio for tasks in the information augmentation scheme with uncertainty is also solved via MDP.

\textbf{State}: Because the correctness of the inference result is unknown, the uncertainty of the inference result should be added into state for decision making, and thus the state of the MDP is changed to $s=\{a,r_L,r,\mathcal{U}\}$, where $\mathcal{U}$ is the uncertainty of the result of last transmission of the HoL task. For the practical implementation of MDP, uncertainty $\mathcal{U}$ should be quantized. We use uniform quantization for uncertainty.

\textbf{Action}: Still, there are two actions for the MDP, transmit or not transmit, denoted by $r_U$.

\textbf{Reward}: If the action is no transmission, the reward is $0$. If the action is to transmit, the reward of state transition from state $s=\{a,r_L,r,\mathcal{U}\}$ to state $s'=\{a',r,r',\mathcal{U}'\}$ is
\begin{equation}
W_U(s,s') = \rho(r|r_L,\mathcal{U},\mathcal{U}') - \rho(r_L|\mathcal{U}).
\end{equation}
Here, $\rho(r|r_L,\mathcal{U},\mathcal{U}')$ is the accuracy of the learning model with compression ratio $r$, conditioned on the compression ratio of last transmission $r_L$, the uncertainty of the result of last transmission $\mathcal{U}$ and the uncertainty of the newest result $\mathcal{U}$'. $\rho(r_L|\mathcal{U})$ is the accuracy of the learning model with compression ratio $r_L$, conditioned on the uncertainty of the result $\mathcal{U}$. $\rho(r|r_L,\mathcal{U},\mathcal{U}')$ and $\rho(r_L|\mathcal{U})$ can be obtained by performing inference on the validation dataset with optional compression ratios.

\textbf{State transition probability}: The state transition probability depends on the arrival process of tasks and the probability distribution of uncertainty $\mathcal{U}$ of learning model. If the action is no transmission, the state $s=\{a,r_L,r,\mathcal{U}\}$ will transit to state $s'=\{a,r_L,r',\mathcal{U}\}$ with probability $1$, where $r'$ is the next considered compression ratio after $r$. If the action is to transmit, state $s=\{a,r_L,r,\mathcal{U}\}$ will transit to state $s'=\{\lfloor \frac{a}{2^{T(r)}} \rfloor + i\times2^{\tau-T(u)},r,r',\mathcal{U}'\}\ (i \in [0,2^{T(r)}-1] \cap \mathbb{N})$ and the state transition probability is
\begin{equation}\label{eq:retansP}
\mathbf{P}(s,s') = p^{B(i)}(1-p)^{T(r)-B(i)} \times \mathbf{P}_U(r,\mathcal{U}'|r_L,\mathcal{U}),
\end{equation}
where $\mathbf{P}_U(r,\mathcal{U}'|r_L,\mathcal{U})$ is the probability that the model output has uncertainty $\mathcal{U}'$ with compression ratio $r$ conditioned on that the model output of the last transmission with compression ratio $r_L$ has uncertainty $\mathcal{U}$.

The number of states is $S_{UR} = (2^{\tau}-1)U|R|^2$ and the size of action space is $2$, where $U$ is the number of the quantization level of $\mathcal{U}$. The memory usage is the simplified state transition matrix mentioned in last subsection and $\mathbf{P}_U$. As a result, the space complexity is $\mathcal{O}(S_{UR})$ and the time complexity is $\mathcal{O}(S_{UR} \times S \times \frac{1}{\tau})$.

\subsection{Packet Loss-Aware Retransmission Scheme}

The transmission of data samples may fail due to the unreliable wireless channels. If the transmission fails, the edge device can retransmit the data before the deadline. Then we need to design a scheme to decide possible retransmissions of the data samples according to the state of task queue. When the transmission fails, the edge server can keep this task in the task queue and update the queue state. The edge server can still use the online algorithm designed in subsection \ref{sec:online} to select compression ratio for the transmission of this task according to the new queue state. This original retransmission scheme is regarded as the baseline of retransmission scheme.

However, the above algorithm does not consider the packet error and retransmissions. Therefore, we introduce PER and retransmissions into the state transition of MDP to design an evolutionary retransmission scheme. Here, the design of MDP meets the same problem as the information augmentation, where the transmission of different tasks should have the same times of state transition in MDP. In the edge learning system we consider, one time of transmission of data samples needs at least one time slot, so the the edge device can transmit the data sample of one task at most $\tau$ times in this system due to the hard deadline $\tau$. Therefore, we use the same method as information augmentation to design MDP to ensure that different tasks have the same number of state transitions. For every task, the edge server virtually considers $\tau$ times of transmission, including whether transmitting and the compression for transmission.

\textbf{State}: The state of MDP is $s=\{a,r_L,\tau_s\}$, where $a$ is the queue state, $r_L$ is compression ratio of last successful transmission (no packet error) and $\tau_s$ is number of considered transmissions.

\textbf{Action}: The action of MDP is the compression ratio for transmission or no transmission for the HoL task.

\textbf{Reward}: For state $s=\{a,r_L,\tau_s\}$, if the action is to transmit with compression ratio $r$ and the transmission succeeds, the reward is
\begin{equation}
W_R(s,r) = \rho(r)-\rho(r_L),
\end{equation}
otherwise the reward is $0$.

\textbf{State transition probability}: The state transition probability depends on the arrival process of tasks, the accuracy of learning model and the PER. With action $r$, the state $s=\{a,r_L,\tau_s\}$ will transit to state $s'=\{\lfloor \frac{a}{2^{T(r)}} \rfloor + i\times2^{\tau-T(r)},r,\tau_s+1\}$ if transmission succeeds, or state $s''=\{\lfloor \frac{a}{2^{T(r)}} \rfloor + i\times2^{\tau-T(r)},r_L,\tau_s+1\}$ if transmission fails. The state transition probability is
\begin{equation}\label{eq:retansP}
\mathbf{P}(s,s') = p^{B(i)}(1-p)^{T(r)-B(i)} \times \left(1-P_e(T(r))\right),
\end{equation}
and
\begin{equation}\label{eq:retansP}
\mathbf{P}(s,s'') = p^{B(i)}(1-p)^{T(r)-B(i)} \times P_e(T(r)).
\end{equation}

For this proposed algorithm, the space complexity is $\mathcal{O}(S \times |R| \times \tau)$ and the time complexity is $\mathcal{O}(S \times S \times |R|)$.

\section{Experiment}\label{sec:exp}

We use two datasets, namely MNIST and cifar10 \cite{cifar10} to evaluate our proposed algorithms. The experiments are based on simulations and the performance of the proposed scheme is evaluated by the proportion of successfully completed tasks.

\subsection{MNIST Dataset}

\subsubsection{Experiment Setup}

MNIST is the handwritten digit images dataset and the task of the edge learning system is the number recognition of an image of MNIST testing dataset. In MNIST, there are 60000 images in the training dataset and 10000 images in the testing dataset. Each image in MNIST has $28 \times 28$ pixels, and corresponds to an integer number from 0 to 9. Training dataset is used for training a machine learning model deployed on the edge server. A task is successfully completed only when it is completed before the deadline \emph{and} its inference result is correct.

The machine learning model deployed on the edge server is a multi-layer perceptron (MLP) \cite{893399} with one hidden layer and 700 hidden units. The output of the model is vector ${\bm X}$ with $10$ elements and $\sum\limits_{i=1}^{10} X_i=1$. The $i$-th element of ${\bm X}$ represents the probability that the input image belongs to class $i$ (corresponding to the number $i-1$).

The compression algorithm of images in the experiment is downsampling. The optional resolution of images for transmission include $4 \times 4$, $7 \times 7$ and $14 \times 14$, for which the compression ratio $r$ is $49$, $16$, $4$. The time of one time slot is normalized to be the time used for transmitting an image with resolution $4 \times 4$. Hence the number of time slots used for transmitting data with these optional resolution are $1$, $3$ and $10$. We use the training dataset to train the model used for performing inference task and obtain the accuracy of the model with respect to different compression ratios, i.e., $\rho(r)$, being $0.89$, $0.97$ and $0.98$, for $r=49, 16, 4$, respectively. The data used for simulations is from the test dataset.

\subsubsection{Experiment Result}

\begin{figure}
    \centering
    \includegraphics[width=0.9\linewidth]{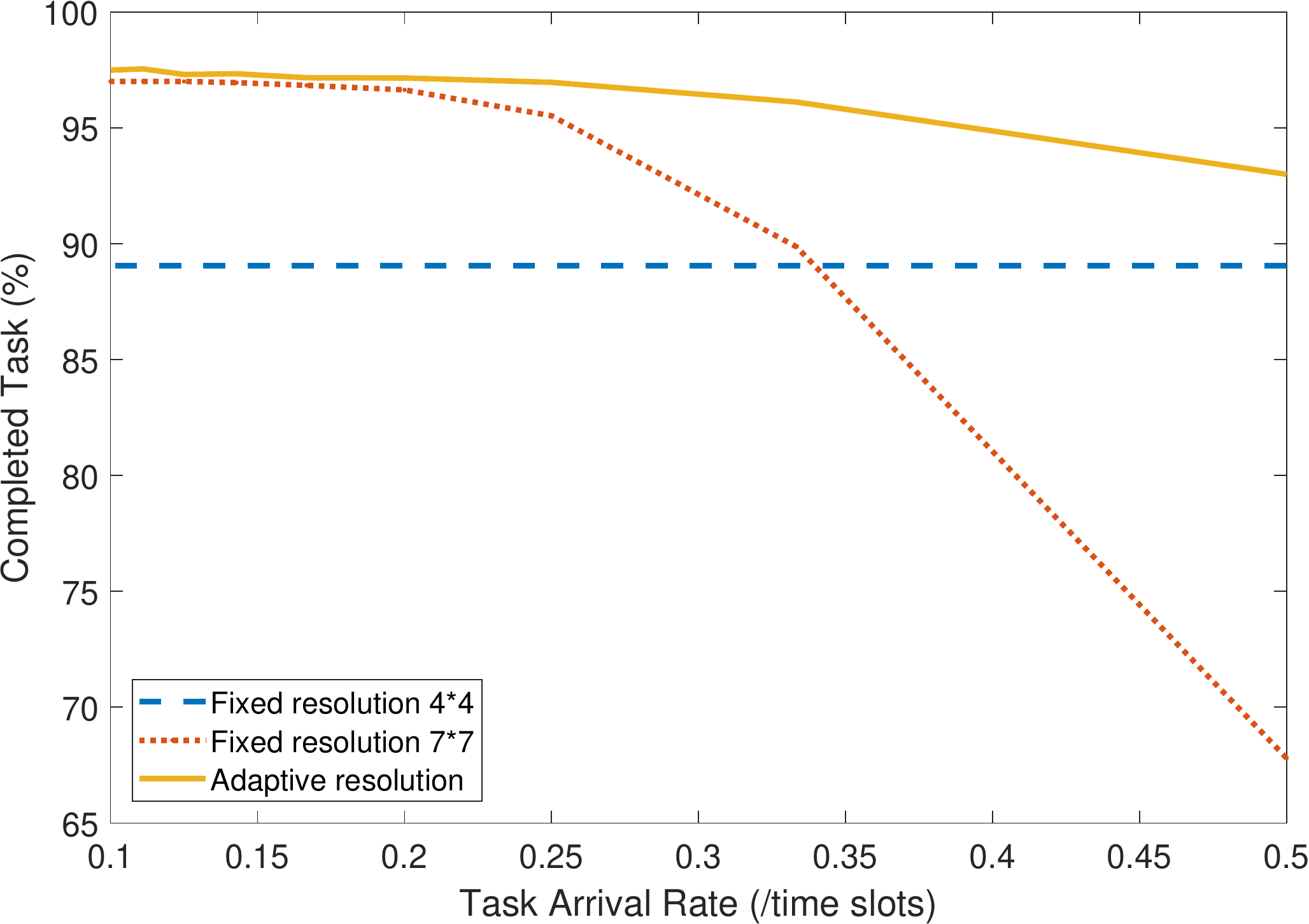}
    \caption{Performance of the proposed offline algorithm with dynamic compression ratio selection under different arrival rates (MNIST dataset).}
    \label{fig:offline}
\end{figure}

\begin{figure}
    \centering
    \includegraphics[width=0.9\linewidth]{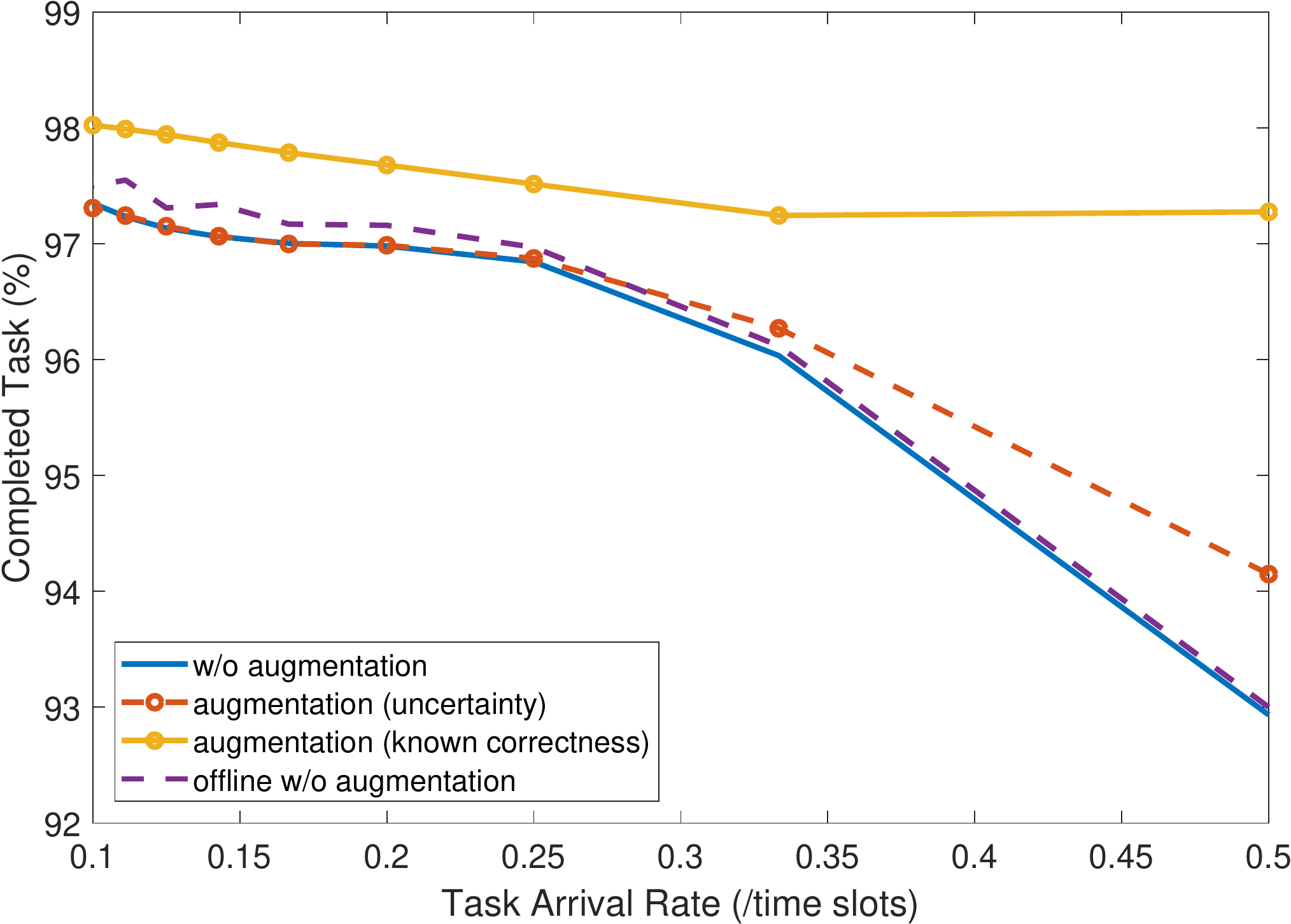}
    \caption{Performance of the proposed online algorithms with dynamic compression ratio selection under different arrival rates (MNIST dataset).}
    \label{fig:online}
\end{figure}

Firstly, we provide the performance versus different task arrival rates. We first suppose that there is no packet loss, which means $p_e = 0$. The latency requirement is $\tau=12$ slots. The number of quantization level of $\mathcal{U}$ of information augmentation scheme is 10. To show the gain of dynamic compression ratio selection, we compare the offline algorithm with the algorithm with fixed resolution. Fig. \ref{fig:offline} shows that the dynamic compression ratio selection can substantially increase the number of completed tasks. The performance of the offline algorithm is the upper bound of the online algorithm, which can be used to evaluate the performance of the proposed online algorithm. The performance of the online algorithm and information augmentation using MDP is shown in Fig. \ref{fig:online}. It is shown that online algorithm has almost the same performance as the offline algorithm. The information augmentation brings improvement, especially when the arrival rate is high. Specifically, when the task arrival rate is $0.5$ per time slot, $97.5\%$ of all tasks can be completed, with $4.5\%$ improvement compared with the algorithm without information augmentation. The proposed information augmentation scheme is also more robust to the changes of the arrival rate. The performance of information augmentation scheme with known inference correctness can be regarded as the upper bound of the information augmentation scheme. Using uncertainty for confidence estimation, the information augmentation scheme can still perform better than the online algorithm without information augmentation. Note that when the arrival rate is low, the improvement of the information augmentation is marginal. It is because tasks can be transmitted with low compression ratio for the first transmission when the arrival rate is low and the information augmentation is unnecessary for many tasks.

\begin{figure}[ht]
    \centering
    \includegraphics[width=0.9\linewidth]{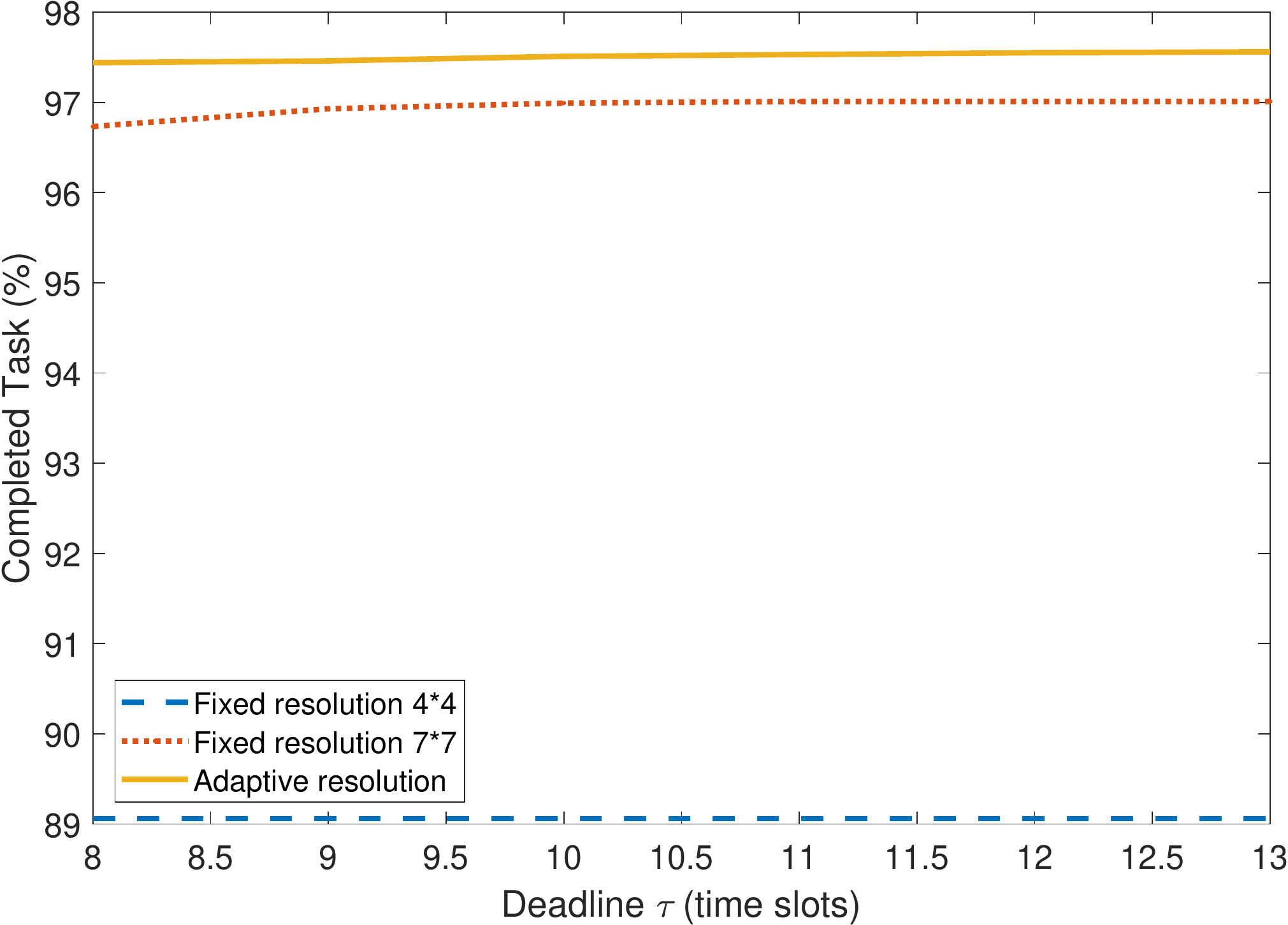}
    \caption{Performance of the proposed offline algorithm with dynamic compression ratio selection under different deadlines (MNIST dataset).}
    \label{fig:offline_tau}
\end{figure}

\begin{figure}[ht]
    \centering
    \includegraphics[width=0.9\linewidth]{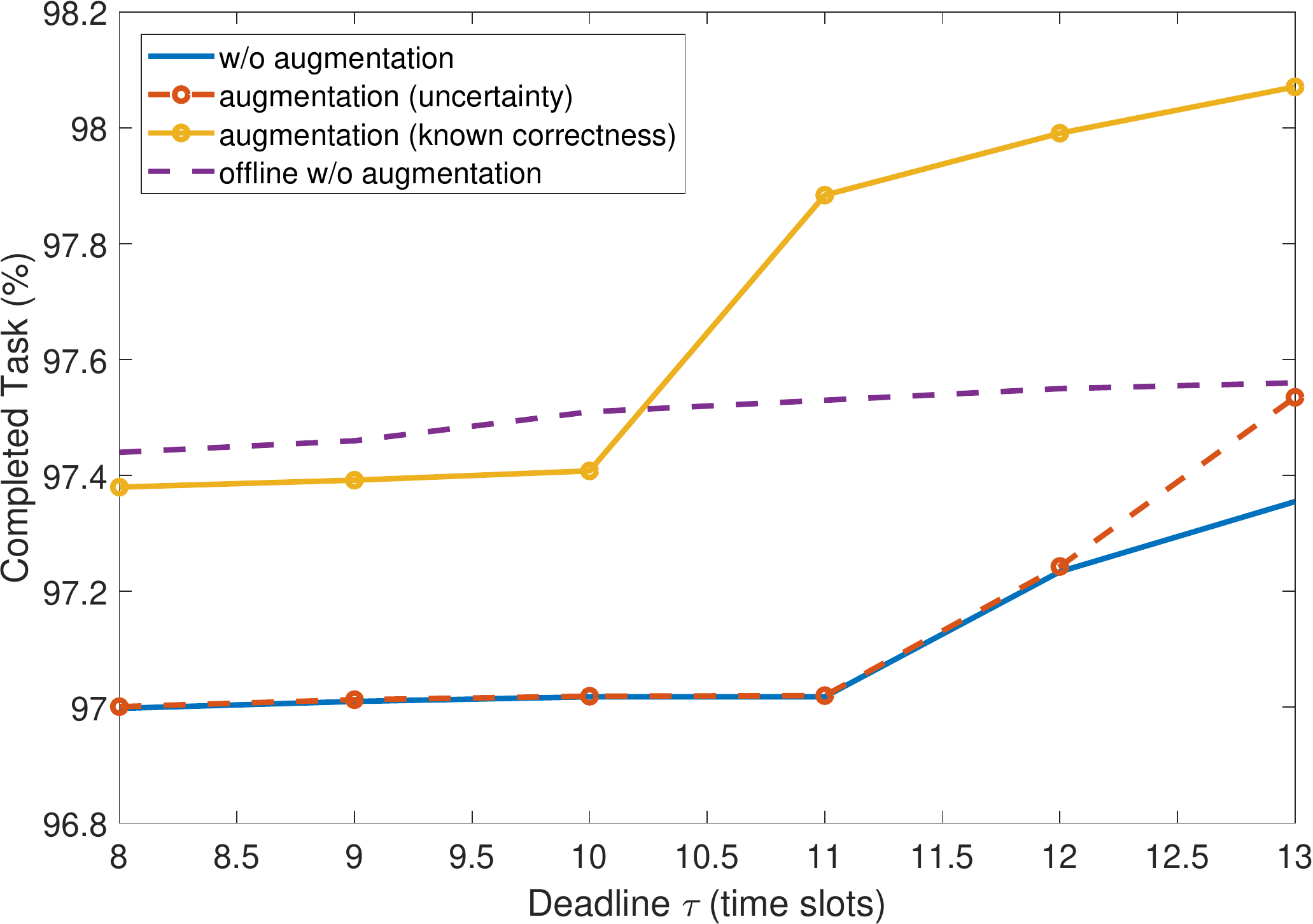}
    \caption{Performance of the proposed online algorithms with dynamic compression ratio selection under different deadlines (MNIST dataset).}
    \label{fig:online_tau}
\end{figure}

Then we provide the performance with different latency requirements. The arrival rate is set to $p=0.11$. $p_e = 0$. Fig. \ref{fig:offline_tau} shows the gain of the offline algorithm with dynamic compression ratio selection, compared with the algorithm with the fixed resolution. Fig. \ref{fig:online_tau} compares the performance of the offline algorithm, online algorithm and information augmentation scheme. The gaps between offline algorithm and online algorithm becomes small when $\tau$ becomes larger. The information augmentation scheme brings more improvement with larger $\tau$. It is because that the information augmentation scheme can reduce the average communication cost but results in extra latency of some tasks, leading to possible task failures. With larger $\tau$, fewer tasks fail and the advantage of information augmentation scheme, that reduces the average communication cost, becomes more significant.

\begin{figure}[ht]
    \centering
    \includegraphics[width=0.9\linewidth]{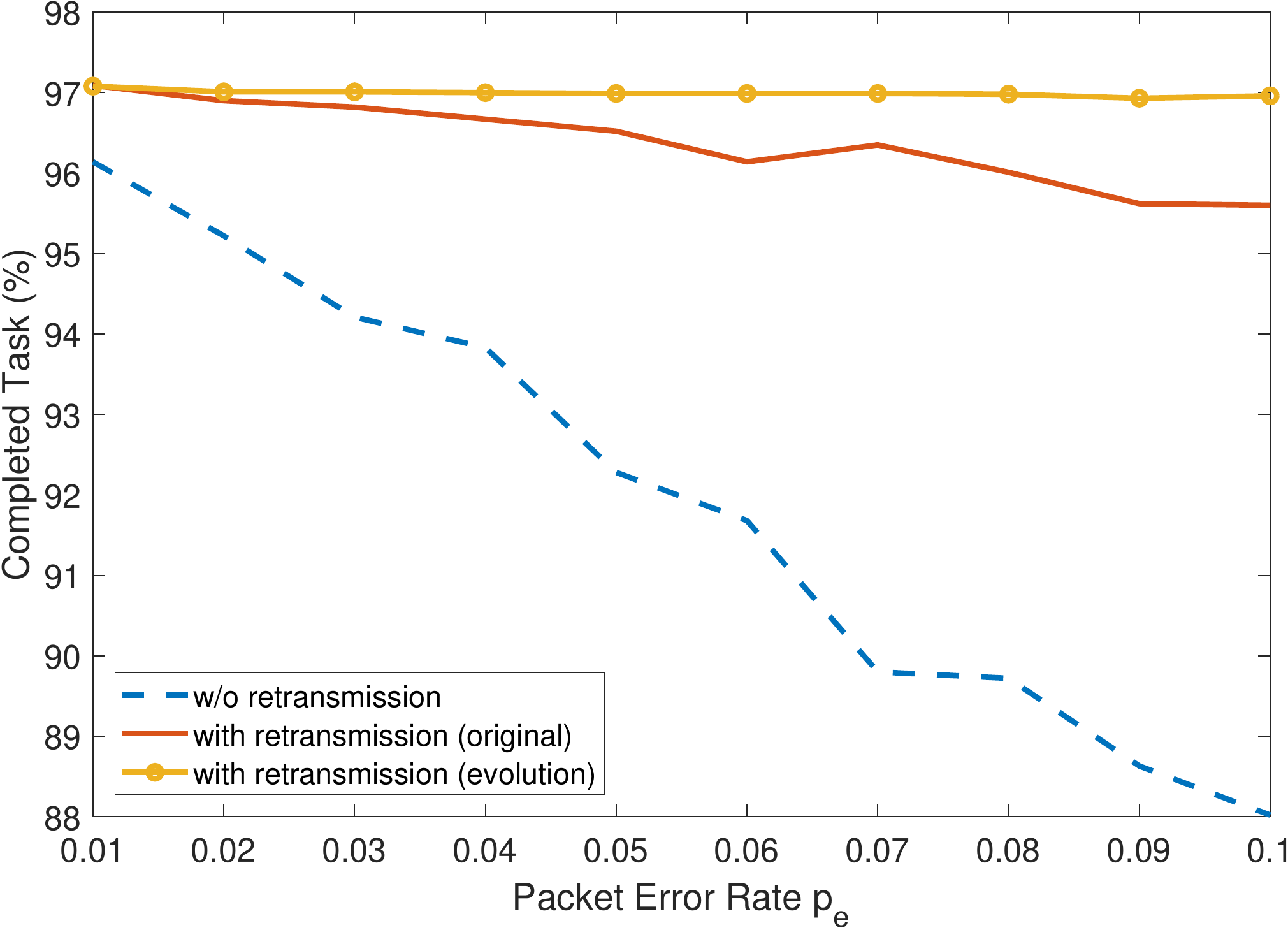}
    \caption{Performance of the proposed online retransmission schemes without information augmentation (MNIST dataset).}
    \label{fig:channel}
\end{figure}

Considering the packet loss in the edge learning system, we compare the performance of the retransmission scheme under different PER. Fig. \ref{fig:channel} shows the performance of three schemes (without information augmentation), including scheme without retransmission, baseline of retransmission scheme (original MDP without considering the packet loss) and the retransmission scheme using MDP that considers packet loss and retransmission. The performance of the scheme without retransmission shows that the packet loss brings severe performance degradation. The retransmission scheme using MDP that considers retransmission when training is much more robust to different PER and improves the performance for over $1\%$ as compared to the baseline when $p_e=0.1$.

\subsection{Cifar10 Dataset}

\subsubsection{Experiment Setup}

Cifar10 is an images dataset with 10 classes (cat, dog, etc.) and the task of the edge learning system is image recognition. In cifar10, there are 50000 images in the training dataset and 10000 images in the testing dataset. Each image in cifar10 has $32 \times 32$ pixels.

The machine learning model deployed on the edge server is mobilenet-v2 with 3.4M parameters \cite{sandler2018mobilenetv2}. The output of the model is vector ${\bm X}$ with $10$ elements and $\sum\limits_{i=1}^{10} X_i=1$. The $i$-th element of ${\bm X}$ represents the probability that the input image belongs to class $i$.

The compression algorithm of images in the experiment is downsampling. The optional resolutions of images for transmission include $12 \times 12$, $16 \times 16$ and $32 \times 32$, for which the compression ratio $r$ is $7$, $4$, $1$. The time of one time slot is normalized to be the time used for transmitting an image with resolution $12 \times 12$. Hence the number of time slots used for transmitting data with these optional resolutions are $1$, $2$ and $8$. We use the training dataset to train the model and obtain the accuracy of the model by performing inference on the training dataset, i.e., $\rho(r)=0.81$, $0.87$ and $0.92$, for compression ratio $r=7, 4, 1$, respectively. The data used for simulations is from the test dataset.

\subsubsection{Experiment Results}

\begin{figure}
    \centering
    \includegraphics[width=0.9\linewidth]{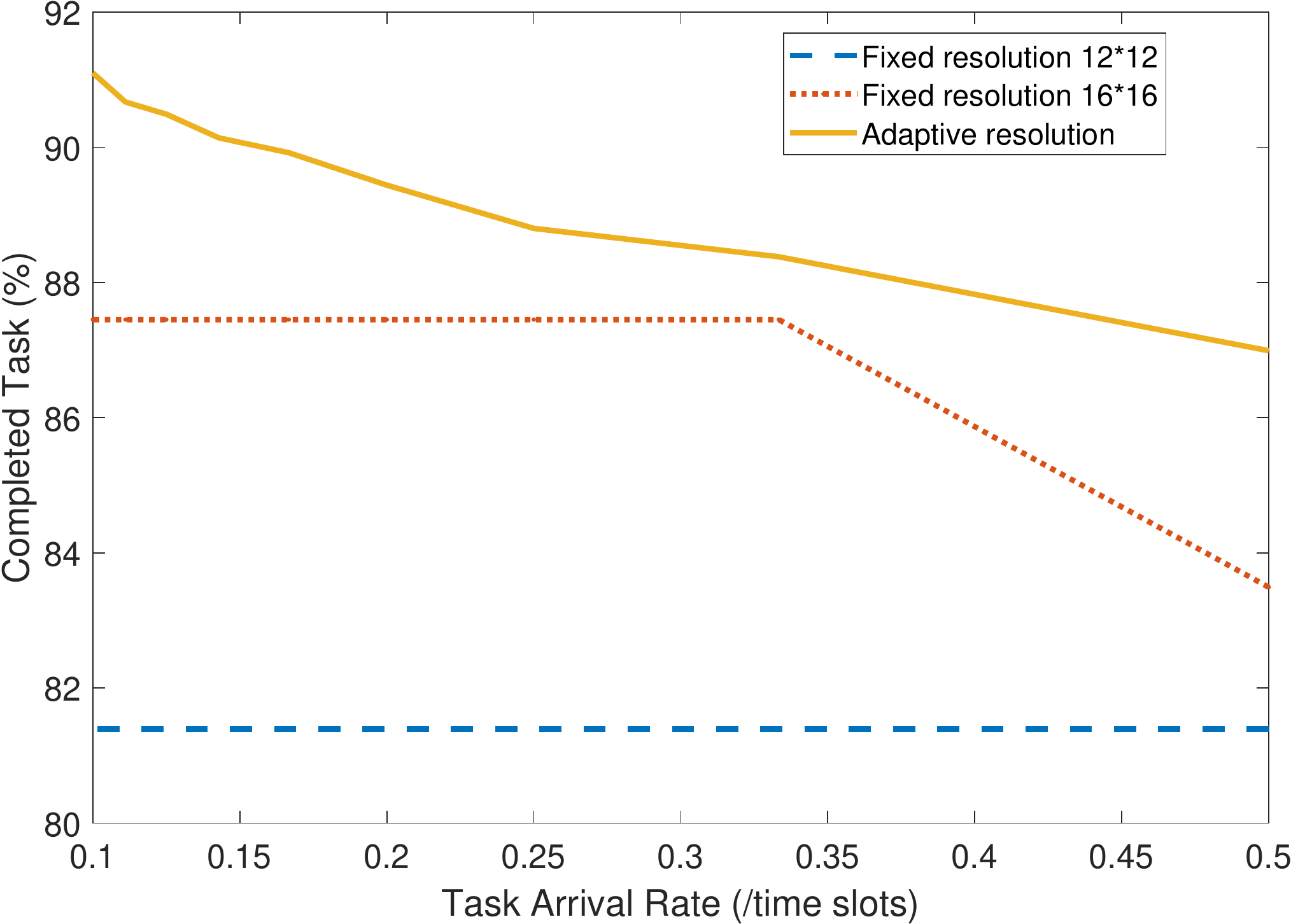}
    \caption{Performance of the proposed offline algorithm with dynamic compression ratio selection under different arrival rates (cifar10 dataset).}
    \label{fig:offline_ci}
\end{figure}

\begin{figure}
    \centering
    \includegraphics[width=0.9\linewidth]{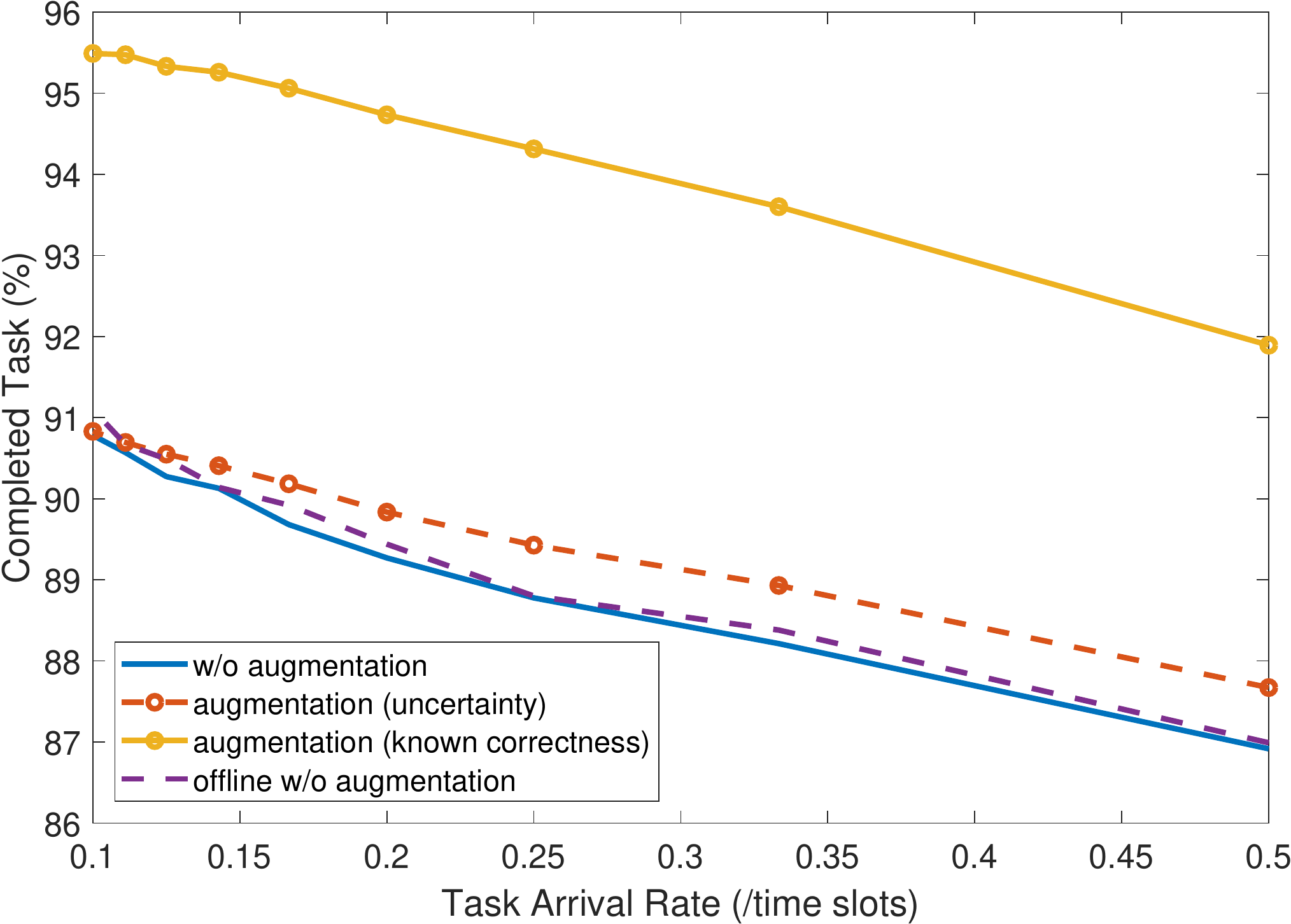}
    \caption{Performance of the proposed online algorithms with dynamic compression ratio selection under different arrival rates (cifar10 dataset).}
    \label{fig:online_ci}
\end{figure}

Firstly, we provide the performance versus different task arrival rates and assume that there is no packet loss ($p_e = 0$). The latency requirement is $\tau=12$ slots. The number of quantization level of $\mathcal{U}$ of information augmentation scheme is also 10. The performance of the offline algorithm of dynamic compression ratio selection is compared with the algorithm that transmits data with fixed resolutions in Fig. \ref{fig:offline_ci}. For the cifar10 dataset, the edge learning system can still substantially increase the number of completed tasks with dynamic compression ratio selection. Then we compare the performance of offline algorithm, online algorithm and information augmentation in Fig. \ref{fig:online_ci}. It is shown that online algorithm has almost the same performance as the offline algorithm (the gap is less than $0.2\%$). The information augmentation (with known correctness) brings about $5\%$ improvement compared with the online algorithm without information augmentation. Using uncertainty for confidence estimation, the improvement resulted by information augmentation becomes less but still brings $1\%$ improvement when the arrival rate is $0.5$. The gain of information augmentation becomes marginal when the arrival rate is lower, which is the same as the result of experiments on MNIST.

\begin{figure}
    \centering
    \includegraphics[width=0.9\linewidth]{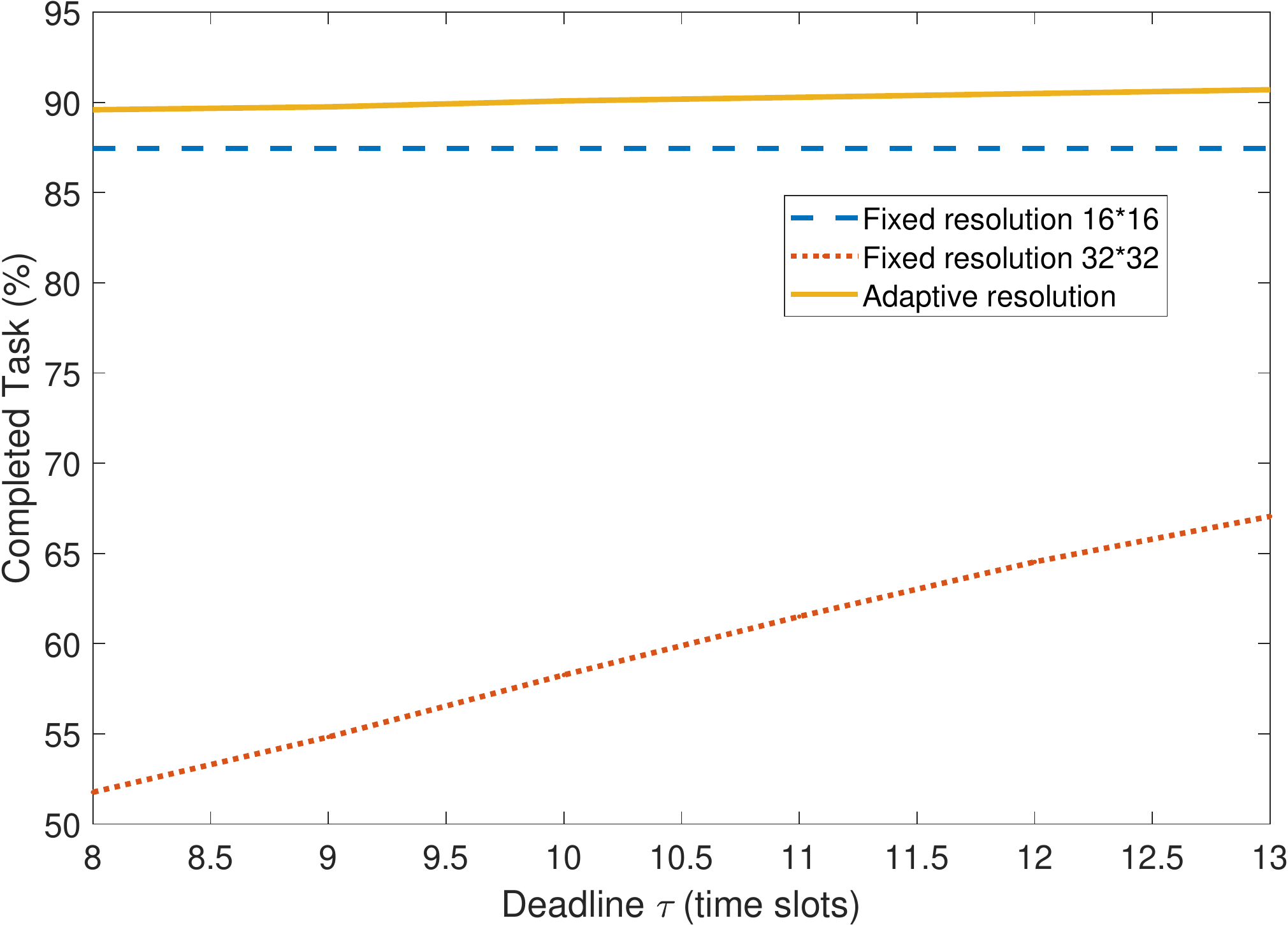}
    \caption{Performance of the proposed offline algorithm with dynamic compression ratio selection under different deadlines (cifar10 dataset).}
    \label{fig:offline_tau_ci}
\end{figure}

\begin{figure}
    \centering
    \includegraphics[width=0.9\linewidth]{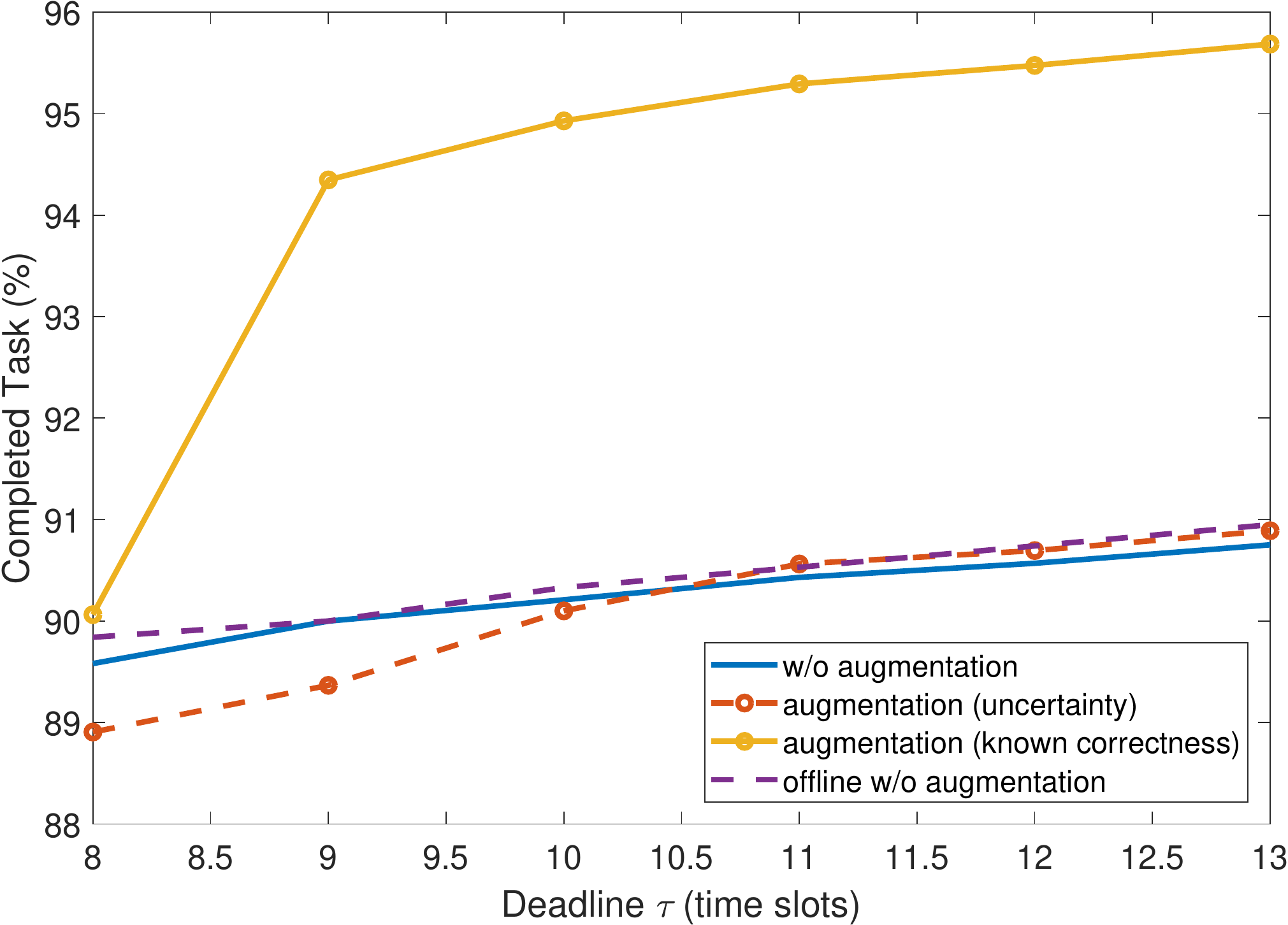}
    \caption{Performance of the proposed online algorithms with dynamic compression ratio selection under different deadlines (cifar10 dataset).}
    \label{fig:online_tau_ci}
\end{figure}

Then we provide the performance with different latency requirements. The arrival rate is $p=0.11$. $p_e = 0$. Fig. \ref{fig:offline_tau_ci} shows the gain of the offline algorithm dynamic compression ratio selection, compared with the fixed resolution counterpart. Fig. \ref{fig:online_tau_ci} compares the performance of offline algorithm, online algorithm and information augmentation scheme. The performance of the online algorithm is almost the same as the offline algorithm. The information augmentation scheme brings more improvement with larger $\tau$, which is the same as the result shown by the experiments on MNIST. When $\tau$ is small, the information augmentation scheme using uncertainty for confidence estimation performs worse than online algorithm without information augmentation. It is because the result of correctness estimation using uncertainty may be wrong, which brings performance degradation. When $\tau$ is small, the performance degradation of uncertainty based confidence estimation is more significant than the improvement of information augmentation, which makes the information augmentation scheme worse than the online scheme without information augmentation.

\begin{figure}
    \centering
    \includegraphics[width=0.9\linewidth]{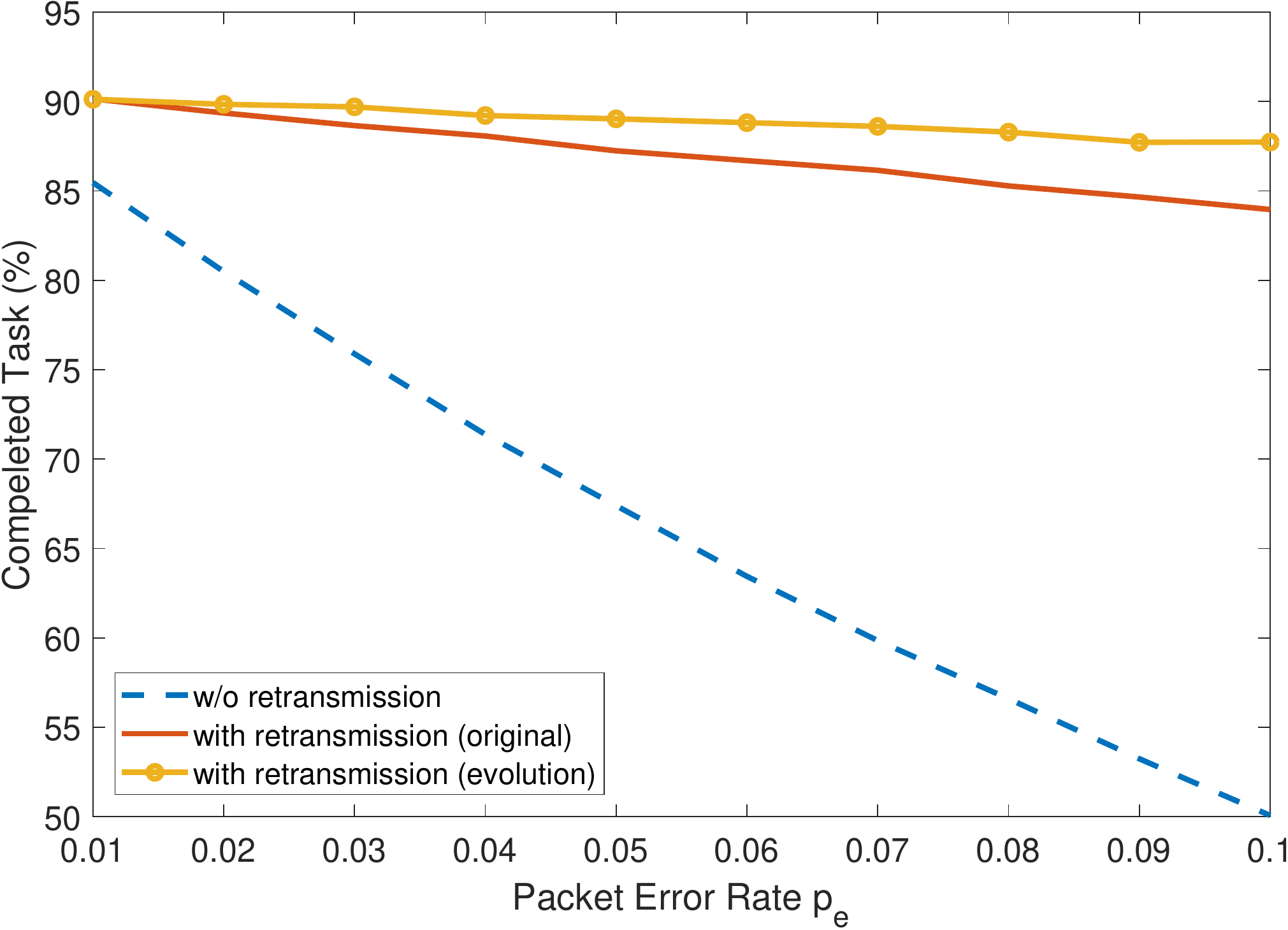}
    \caption{Performance of the proposed online retransmission schemes without information augmentation (cifar10 dataset).}
    \label{fig:channel_ci}
\end{figure}

At last, we do experiments to compare the performance of the retransmission scheme with different PERs. Fig. \ref{fig:channel_ci} shows the performance of three schemes: scheme without retransmission, baseline of retransmission scheme (original MDP) and the retransmission scheme using MDP that considers packet loss and retransmissions. Both retransmission schemes brings notable improvement compared with the scheme without retransmission. The retransmission scheme using MDP that considers retransmission for training is much more robust to the change of PER and performs better than the baseline with about $4\%$ improvement when $p_e=0.1$.

\section{Conclusion}\label{sec:conclusion}

In this paper, we have exploited lossy compression to reduce the transmission latency for an edge inference system with hard deadlines. MDP is utilized to find the optimal dynamic compression ratio selection algorithm, so as to balance the tradeoff between the transmission latency and inference accuracy. Experiments show that dynamic compression ratio selection helps edge inference system complete more tasks under the latency deadline constraint and the proposed online algorithm can get almost the same performance as the offline upper bound. Furthermore, we propose an information augmentation scheme to reduce the communication cost. The information augmentation scheme is more robust to the change of the arrival rate than the one without information augmentation and its performance becomes better when the latency requirement $\tau$ is larger. Uncertainty based confidence estimation for the inference result enables the system to use information augmentation scheme when the correctness of inference result unknown, and about $2\%$ more tasks can be completed when the arrival rate is high, compared with the online algorithm without information augmentation. Finally, the proposed retransmission scheme can further address the unreliable wireless channel, the performance is robust even when the PER becomes larger in experiments. To extend our work in IoT systems with large amount of edge devices, where multiple edge servers should be deployed, the algorithms coordinating task allocation among edge servers are worth studying. The joint inference scheduling and model training is also a promising future direction for many IoT systems.

\bibliographystyle{IEEEtran}
\bibliography{ref}

\end{document}